\documentclass[notoc]{JHEP3}
\pdfoutput=1 
\usepackage{graphicx}
\usepackage{amsmath}
\usepackage{amsfonts}
\usepackage{amssymb}
\usepackage{slashed}

\newcommand{\Mvariable}[1]{}
\newcommand{\GeV}{\,{\rm{GeV}}}

 \providecommand{\GeV}{{\rm{GeV}}}
 
 \providecommand{\imag}[1]{\,i\,}

\title{A short comparison between $m_{T2}$ and $m_{CT}$}

\author{Mario Serna \\ \small
Rudolf Peierls Centre for Theoretical Physics,
University of Oxford, 1 Keble Road, Oxford, OX1 3NP
\\
Email:  \email{mariojr@alum.mit.edu}
 \normalsize
 }

\abstract{
We compare $m_{T2}$ with $m_{CT}$; both are kinematic variables designed to find relationships between masses of pair-produced new states with symmetric decay chains.  We find that for massless visible particles $m_{CT}$ equals  $m_{T2}$ in a particular limit.  We identify advantages and disadvantages to the use of each variable.  Tovey's paper on $m_{CT}$ also introduced a powerful concept of extracting mass information from an analysis at intermediate stages of a symmetric decay chain.  We suggest that $m_{T2}$ is a better tool for performing this analysis than $m_{CT}$ due to $m_{T2}$'s better properties under initial state radiation.
}


\begin{document}


Dark matter's likely signature in a hadron collider will be missing transverse momentum.  The stability of dark matter suggests a charge or conservation law that requires  dark matter particles be produced in pairs at colliders.  The kinematic variables $m_{T2}$ introduced by Lester and Summers
\cite{Lester:1999tx}and $m_{CT}$
introduced by Tovey \cite{Tovey:2008ui} aid in the task of determining the mass of new states that decay to dark matter particles at hadron colliders. Although $m_{T2}$ has been used extensively (see \cite{Barr:2003rg,Lester:2007fq,Cho:2007qv,Cho:2007dh,Ross:2007rm} for a few examples), the variable $m_{CT}$ is new but shares many similarities and differences with $m_{T2}$. This note briefly defines $m_{T2}$ and $m_{CT}$, explains when they give identical results, when they differ, and comments on benefits of each in their intended applications.

Both variables assume a pair-produced new-particle state followed by each branch decaying symmetrically to visible states and dark-matter candidates which escape detection and appear as missing transverse momentum. Fig \ref{FigEventTopology} is the simplest example on-which we can meaningfully compare the two kinematic quantities.  The figure shows two partons colliding and producing some observed initial state radiation (ISR) with four momenta $g$ and an on-shell, pair-produced new state $Y$.  On each branch, $Y$ decays to on-shell states $X$ and $v_1$ with masses $m_{X}$ and $m_{v_1}$, and $X$ then decays to on-shell states $N$ and $v_2$ with masses $m_N$ and $m_{v_2}$.  The  four-momenta of $v_1$, $v_2$ and $N$ are respectively $\alpha_1$, $\alpha_2$ and $p$ on one branch and $\beta_1$, $\beta_2$ and $q$ in the other branch.  The missing transverse momenta $\slashed{P}_T$ is given by the transverse part of $p+q$.
\FIGURE[b]{
\centerline{\includegraphics[width=3in]{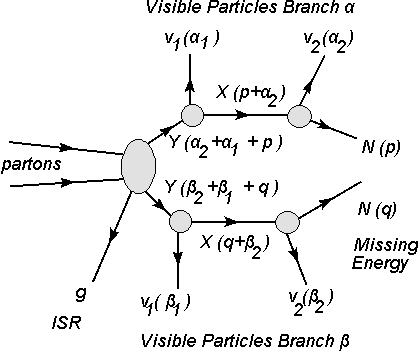}}
\caption{\label{FigEventTopology} A simplest topology with which we compare $m_{T2}$ and $m_{CT}$.}%
}

First we describe $m_{T2}$.
The variable $m_{T2}$ accepts three inputs: $\chi$ (an assumed mass of the two particles carrying away missing transverse momenta), $\alpha$ and $\beta$ (the visible momenta of each branch), and $\slashed{P}_T=(p+q)_T$ (the missing transverse momenta).  The variable $m_{T2}$ is the minimum mass of the pair of parent particles compatible with the observed particles' four momenta and an assumed mass for particles carrying away the missing momenta.  We can define $m_{T2}$ in terms of the transverse mass of each branch where we minimize the maximum of the two transverse masses over the unknown split between $p$ and $q$ of the overall missing transverse momenta:
 \begin{equation}
  m^2_{T2}(\chi_N,\alpha,\beta,\slashed{P}_T) = \min_{{p}_T+{q}_T=\slashed{{P}}_T} \left[ \max \left\{ m^2_T(\alpha,p), m^2_T(\beta,q) \right\} \right].
 \end{equation}
In this expression $\chi_N$ is the assumed mass of $N$, $\alpha$ and $\beta$ are the four momenta of the visible particles in the two branches, the transverse mass is given by $m^2_T(\alpha,p) = m^2_\alpha + \chi_N^2 + 2 (E_T(p) E_T(\alpha) - p_T \cdot \alpha_T)$ and the transverse energy $E_T(p) = \sqrt{p_T^2 + \chi_N^2}$ is determined from the transverse momentum of $p$ and the assumed mass of the particle associated with momentum $p$. An analytic formula for the case with no transverse ISR can be found in the appendix of \cite{Lester:2007fq}.   For each event, the quantity $m_{T2}(\chi_N=m_N,\alpha_1+\alpha_2, \beta_1 + \beta_2, \slashed{P}_T)$ gives the smallest mass for the parent particle compatible with the event's kinematics.  Under ideal assumptions, the mass of the parent particle $Y$ is given by the end-point of the distribution of this $m_{T2}$ parameter over a large number of events like fig \ref{FigEventTopology}.  Because a priori we do not know $m_N$, we need some other mechanism to determine $m_N$ \footnote{The true $m_N$ and $m_Y$ can be found in the case where $Y$ undergoes a three-body ($X$ is off-shell) through kinks in $m_{T2}$ \cite{Cho:2007qv,Cho:2007dh,Barr:2007hy} or when combined with endpoints from other distribution (like $\max m_{ll}$) \cite{Ross:2007rm}. }.  We use $\chi$ to distinguish assumed values of the masses ($\chi_Y$, $\chi_X$, $\chi_N$) from the true values for the masses ($m_Y$, $m_X$, $m_N$).   Because of this dependence on the unknown mass, we should think of ${max}\, m_{T2}$ as providing a relationship or constraint between the mass of $Y$ and the mass of $N$.  This forms a surface in the ($\chi_Y$, $\chi_X$, $\chi_N$) space on which the true mass will lie.  We express this relationship as $\chi_Y(\chi_N)$ \footnote{In principle this surface would be considered a function of $\chi_Y(\chi_X,\chi_N)$, but $m_{T2}$ makes no reference to the mass of $X$ and the resulting constraints are therefore independent of any assumed value of the mass of $X$.}.

Tovey \cite{Tovey:2008ui} recently defined a new variable $m_{CT}$ which has many similarities to $m_{T2}$.  The variable is defined as
 \begin{eqnarray}
   m^2_{CT}(\alpha_1,\beta_1) & = & (E_T(\alpha_1)+E_T(\beta_1))^2 - (\alpha_{1T}-{\beta}_{1T})^2.
 \end{eqnarray}
Tovey's goal is to identify another constraint between masses in the decay chain.  He observes that in the rest frame of $Y$ the momentum of the back-to-back decay products $X$ and $v_1$ is given by
 \begin{equation}
   \left(k_*(m_Y,m_X,m_{v_1}) \right)^2 = \frac{(m_Y^2 - (m_{v_1}+m_X)^2)(m_Y^2 - (m_{v_1}-m_X)^2)}{4 m_Y^2} \label{Eq2BEMP}
 \end{equation}
where $k_*$ is the two-body excess momentum parameter (2BEMP) \footnote{Tovey refers to this as the 2-body mass parameter ${\mathcal{M}}_i$.  We feel calling this a mass is a bit misleading so we are suggesting 2BEMP.}.  In the absence of transverse ISR ($g_T=0$) and if the visible particles are effectively massless ($m_{v_1}=0$), Tovey observes that $\max m_{CT}(\alpha_1,\beta_1)$ is given by $2 k_*$; this provides an equation of constraint between $m_Y$ and $m_X$.  Tovey observes that if we could do this analysis at various stages along the symmetric decay chain all the masses could be determined.

The big advantage of $m_{CT}$ is in its computational simplicity.  Also, $m_{CT}$ is intended to only be calculated once per event instead of at a variety of choices of $\chi$.  In contrast, $m_{T2}$ is a more computationally intensive parameter to compute; but this is aided by the use of a shared repository of community tested C++ libraries found at \cite{AtlasMT2Wiki}.

How are these two variables similar?  Both $m_{CT}$ and $m_{T2}$, in the absences of ISR, are invariant under back-to-back boosts of the parent particles momenta \cite{Cho:2007dh}.  The variable $m_{CT}$ equals $m_{T2} (\chi=0)$ in the special case where $\chi=0$ and when the visible particles are massless $(\alpha_1^2=\beta_1^2=0)$ and the there is no transverse ISR ($g_T=0$)
 \begin{eqnarray}
   m_{CT}(\alpha_1, \beta_1) & = & m_{T2}(\chi=0,\alpha_1,\beta_1, \slashed{P}_T = (p+q+\alpha_2 +\beta_2)_T) \ \ \ {\rm{if}} \ \ \alpha_1^2=\beta_1^2=0. \label{EqMCTMT2Equality} \\
    & = & 2 ( {\alpha_1}_T \cdot {\beta_1}_T + |{\alpha_1}_T |\,| {\beta_1}_T |).  \label{EqMCTMT2Equality2}
 \end{eqnarray}
The $m_{CT}$ side of the equation is straight forward. The $m_{T2}$ side of the expression can be derived analytically using the formula for $m_{T2}$ given in \cite{Lester:2007fq}; we also outline a short proof in the appendix. Eq(\ref{EqMCTMT2Equality}) uses a $m_{T2}$ in a unconventional way;  we group the observed momenta of the second decay products into the missing transverse momenta.    In this limit, both share an endpoint of $2 k_* = (m_Y^2 - m_X^2)/m_Y$. To the best of our knowledge, this endpoint was first pointed out by Cho \emph{et.al.} \cite{Cho:2007qv} \footnote{The endpoint given by Cho \emph{et.al.} is violated for non-zero ISR at $\chi_N < m_N$ and $\chi_N > m_N$.}.  We find it surprising that a physical relationship between the masses follows from $m_{T2}$ evaluated at a non physical $\chi$.  In the presence of ISR, eq(\ref{EqMCTMT2Equality}) is no longer an equality.  Furthermore in the presence of the ISR, the end point of the distribution given by either side of eq(\ref{EqMCTMT2Equality}) exceeds $2 k_*$.  In both cases, we will need to solve a combinatoric problem of matching visible particles to their decay order and branch of the event which is beyond the scope of this paper.

In the case where the visible particle $v_1$ is massive, the two parameters give different end-points
 \begin{eqnarray}
   \max m_{CT} (\alpha_1,\beta_1) & = &  \frac{m_Y^2 - m_X^2}{m_Y} + \frac{m_{v_1}^2}{m_{Y}} \label{EqMassiveMCTLimit} \\
   \max m_{T2}(\chi=0,\alpha_1,\beta_1, \slashed{P}_T = (p+q+\alpha_2 +\beta_2)_T) & = &   \sqrt{ m_{v_1}^2 + 2(k_*^2 + k_* \sqrt{k_*^2+m^2_{v_1}})} \label{EqMassiveMT2Limit}
 \end{eqnarray}
where $k_*$ is given by eq(\ref{Eq2BEMP}).  Unfortunately, there is no new information about the masses in these two endpoints.  If we solve eq(\ref{EqMassiveMCTLimit}) for $m_X$ and substitute this into eq(\ref{EqMassiveMT2Limit}) and (\ref{Eq2BEMP}), all dependence on $m_Y$ is eliminated.

Tovey's idea of analyzing the different steps in a symmetric decay chain to extract the masses is powerful.  Up until now, we have been analyzing both variables in terms of the first decay products of $Y$.  This restriction is because $m_{CT}$ requires no transverse ISR to give a meaningful endpoint.  If we were to try and use $\alpha_2$ and $\beta_2$ to find a relationship between $m_X$ and $m_N$, then we would need to consider the transverse ISR to be $(g+ \alpha_1+\beta_1)_T$ which is unlikely to be zero.

\FIGURE{
\centerline{\includegraphics[width=3in]{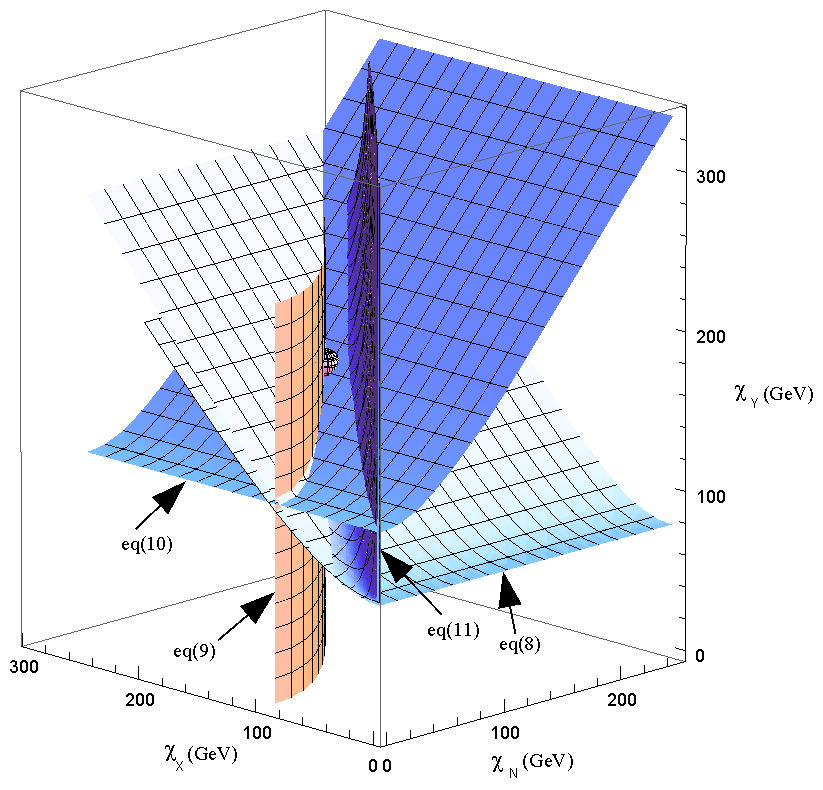}}
\caption{\label{FigCombinedConstraints} Shows constraints from $\max\,m_{T2}$ used with different combinations as described in eqs(\ref{EqYGivenX},\ref{EqXGivenN},\ref{EqYGivenN}) and the $\max m_{12}$ described in eq(\ref{EqmllConstraint}).  Intersection is at the true mass $(97 \GeV,144 \GeV,181 \GeV)$ shown by sphere. Events include ISR but otherwise ideal conditions: no background, resolution, or combinatoric error.}%
}
We suggest $m_{T2}$ is a better variable with which to implement Tovey's idea of analyzing the different steps in a symmetric decay chain because of its ISR properties.
With and without ISR, $m_{T2}$'s endpoint gives the correct mass of the parent particle when we assume the correct value of the missing-energy-particle's mass \footnote{In principle we could plot the $\max m_{T2}(\chi_X,\alpha_1,\beta_1,\slashed{P}_T=(\alpha_2+\beta_2+p+q)_T)$ vs $\chi_X$ as a function of transverse ISR and the value of $\chi_X$ at which the end point is constant would give the correct value of $m_X$; at which point the distributions end point would give the correct $m_Y$.  In practice we probably will not have enough statistics of ISR events.}.
For this reason, $\max m_{T2}$ gives a meaningful relationship between masses $(m_Y,m_X,m_N)$ for all three symmetric pairings of the visible particles across the two branches.  A relationship between $m_Y$ and $m_X$ is given by
 \begin{equation}
 \chi_Y(\chi_X) =  \max m_{T2}(\chi_X,\alpha_1,\beta_1, \slashed{P}_T = (p+q+\alpha_2 +\beta_2)_T). \label{EqYGivenX}
 \end{equation}
A relationship between $m_X$ and $m_N$  can be found by computing
 \begin{equation}
 \chi_X(\chi_N) =  \max m_{T2}(\chi_N,\alpha_2,\beta_2, \slashed{P}_T = (p+q)_T)  \label{EqXGivenN}
 \end{equation}
where we have grouped $\alpha_1+\beta_1$ with the $g$ as a part of the ISR.
A relationship between $m_Y$ and $m_N$ can be found by using $m_{T2}$ in the traditional manner giving
 \begin{equation}
 \chi_Y(\chi_N) =  \max m_{T2}(\chi_N,\alpha_1+\alpha_2,\beta_1+\beta_2, \slashed{P}_T = (p+q)_T).  \label{EqYGivenN}
 \end{equation}
Lastly, we can form a distribution from the invariant mass of the visible particles on each branch $m^2_{12}=(\alpha_1+\alpha_2)^2$ or  $m^2_{12}=(\beta_1+\beta_2)^2$.  The endpoint of this distribution gives a relationship between $m_Y$, $m_X$, and $m_N$ given by
 \begin{equation}
   \max m^2_{12} = \frac{(m^2_Y-m_X^2)(m_X^2-m^2_N)}{m_X^2}.  \label{Eqmll}
 \end{equation}
Solving this expression for $m_Y$ gives the relationship
 \begin{equation}
  \chi^2_Y (\chi_N,\chi_X) = \frac{\chi_X^2 ( (\max m^2_{12}) +\chi_X^2 - \chi_N^2)}{\chi_X^2 - \chi_N^2} \label{EqmllConstraint}.
 \end{equation}
Fig \ref{FigCombinedConstraints} shows the constraints from eqs(\ref{EqYGivenX},\ref{EqXGivenN},\ref{EqYGivenN},\ref{EqmllConstraint}) in an ideal simulation using $(m_Y=181 \, {\rm{GeV}}$, $m_X=144\, {\rm{GeV}}$, $m_N=97\, {\rm{GeV}})$, 1000 events, and massless visible particles, and ISR added with an exponential distribution with a mean of $50$ GeV.  These four surfaces in principle intersect at a single point $(m_Y, m_X, m_N)$ given by the sphere in the figure \ref{FigCombinedConstraints}.  Unfortunately, all these surfaces intersect the correct masses at a shallow angles so we have a sizable uncertainty along the direction of the sum of the masses and a tight constraints in the perpendicular directions. In other words, the mass differences are well-determined but not the mass scale.   From here one could use a shape fitting technique like that described in \cite{Ross:2007rm} to find a constraint on the sum of the masses.  Tovey's suggestion for extracting information from these intermediate stages of a symmetric cascade chain clearly provides more constraints to isolate the true mass than one would find from only using the one constraint of eq(\ref{EqYGivenN}) as described in \cite{Cho:2007qv}.  However, Tovey's suggestion
is more feasible using the $m_{T2}$ rather than $m_{CT}$ because the constraint surfaces derived from $m_{T2}$ intersect the true masses even with ISR.

In summary, we have compared and contrasted $m_{CT}$ with $m_{T2}$.  The variable $m_{CT}$ is a special case of $m_{T2}$ given by eq(\ref{EqMCTMT2Equality}) when ISR can be neglected and when the visible particles are massless.  In this case, the end-point of this distribution gives $2 k_*$, twice the two-body excess momentum parameter (2BEMP). If $m_{v_1} \neq 0$, the two distributions have different endpoints but no new information about the masses.   In the presence of ISR the two functions are not equal; both have endpoints that exceed $2 k_*$.  Because of it's better properties in the presence of ISR, $m_{T2}$ is a better variable for the task of extracting information from each step in the decay chain.  Extracting this information requires solving combinatoric problems which are beyond the scope of this paper.
%

\acknowledgments

MS like to thank Laura Serna, Alan Barr, Chris Lester, Graham Ross, and Dan Tovey for many helpful comments and reviewing the manuscript.
MS would also like to thank Alan Bar and Chris Lester
for suggesting that we write this short comparison between $m_{CT}$ and $m_{T2}$.
MS acknowledges support from the United States Air Force Institute of Technology.
The views expressed in this letter are those of the author and
do not reflect the official policy or position of the United States Air
Force, Department of Defense, or the US Government.

\appendix

\section*{Appendix: Verifying $m_{T2}$ in eq(\ref{EqMCTMT2Equality2})}

We derived the $m_{T2}$ side of eq(\ref{EqMCTMT2Equality2}) by following the analytic solution given by Barr and Lester in \cite{Lester:2007fq}.  In this appendix, we outline how to verify that $m_{T2}$ is is indeed given by
  \begin{equation}
    m_{T2} (\chi=0, \alpha , \beta , \slashed{P}_T = -\alpha_T - \beta_T) = 2 ( {\alpha}_T \cdot {\beta}_T + |{\alpha}_T |\,| {\beta}_T |) \label{EqAppendixEquality}
  \end{equation}
when $\alpha^2=\beta^2=0$ and $p^2=q^2=\chi^2=0$ and $g_T=0$.  To do this we note that $m_{T2}$ can also be defined as the minimum value of $(\alpha+p)^2$ minimized over $p$ and $q$ subject to the conditions $p^2=q^2=\chi^2$ (on-shell missing energy state), and $(\alpha+p)^2 = (\beta+q)^2$ (equal on-shell parent-particle state), and  $(\alpha+\beta+p+q+g)_T=0$ (conservation of transverse momentum) \cite{Ross:2007rm}.

The solution which gives eq(\ref{EqAppendixEquality}) has $p_T = -\beta_T$ and $q_T=-\alpha_T$ with the rapidity of $p$($q$) equal to the rapidity of $\alpha$ ($\beta$).  We now verify that this solution satisfies all the constraints listed above.  Transverse momentum conservation is satisfied trivially: $(\alpha+\beta+p+q)_T=(\alpha+\beta-\alpha -\beta)_T =0$.  The constraint to have the parent particles on-shell can be verified with $2|\alpha_T||p_T| - 2\vec{p}_T  \cdot \vec{\alpha}_T =2|\beta_T||q_T| - 2\vec{q}_T \cdot \vec{\beta}_T  = 2|\beta_T||\beta_T| + 2\vec{\alpha}_T \cdot \vec{\beta}_T$.

Now all that remains is to show that the parent particle's mass is a minimum with respect to ways in which one splits up the missing transverse momentum between $p_T$ and $q_T$ while satisfying the above constraints.  We take $p$ and $q$ to be a small deviation from the stated solution ${p}_T = -{\beta}_T +{\delta}_T$ and ${q}_T = -\alpha_T - {\delta}_T$ where $\delta_T$ is the small deviation in the transverse plane.   We keep $p$ and $q$ on shell at $\chi=0$. The terms $p_o$, $p_z$, $q_o$, $q_z$ are maintained at their minimum by keeping the rapidity of $p$ and $q$ equal to $\alpha$ and $\beta$.  The condition that the parent particles are on-shell and equal is satisfied for a curve of values for $\delta_T$.  The deviation tangent to this curve near $|\delta_T|=0$ is given by
$\delta_T(\lambda) = \lambda\, \hat{z} \times (\alpha_T |\beta_T| + \beta_T |\alpha_T|)$ where $\times$ is a cross product, $\hat{z}$ denotes the beam direction, and we
parameterized the magnitude by the scalar $\lambda$.  Finally, we can substitute $p$ and $q$ with the deviation $\delta_T(\lambda)$ back into the expression for the parent particle's mass $(\alpha+p)^2$  and verify that $2 ( {\alpha}_T \cdot {\beta}_T + |{\alpha}_T |\,| {\beta}_T |)$ at $\lambda=0$ is indeed the minimum with respect to changes in $\lambda$.


\newcommand{\noopsort}[1]{} \newcommand{\printfirst}[2]{#1}
  \newcommand{\singleletter}[1]{#1} \newcommand{\switchargs}[2]{#2#1}
\providecommand{\href}[2]{#2}\begingroup\raggedright\endgroup

\end{document}